\documentclass{elsart3} 
\usepackage{graphicx}% Include figure files
\usepackage{dcolumn}% Align table columns on decimal point
\usepackage{bm}% bold math
\usepackage{natbib}
\usepackage{amssymb}
\usepackage{amsmath}
\usepackage{txfonts}
\begin{document}
\begin{frontmatter}

%\preprint{Physical Review B, In Press}

\title{STM observation of the quantum interference effect\\in finite-sized graphite}% Force line breaks with \\

\author[label1]{Yousuke Kobayashi}
\ead{ykobaya@chem.titech.ac.jp}
%\author[label1]{Kazuyuki Takai}%less active
\author[label1]{Ken-ichi Fukui}
\author[label1]{Toshiaki Enoki}
\author[label2,label3]{Kikuo Harigaya}
\author[label4]{Yutaka Kaburagi}
\author[label4]{Yoshihiro Hishiyama}

\address[label1]{Department of Chemistry, Tokyo Institute of Technology, 2-12-1, Ookayama, Meguro-ku, Tokyo 152-8551, Japan}
\address[label2]{Nanotechnology Research Institute, AIST, Tsukuba 305-8568, Japan}
\address[label3]{Synthetic Nano-Function Materials Project, AIST, Tsukuba 305-8568, Japan}
\address[label4]{Department of Energy Science and Engineering, Musashi Institute of Technology,\\1-28-1, Tamazutsumi, Setagaya-ku, Tokyo 158-8557, Japan}


\begin{abstract}

Superperiodic patterns were observed by STM on two kinds of finite-sized graphene sheets. One is nanographene sheets 
inclined from a highly oriented pyrolitic graphite (HOPG) substrate and the other is several-layer-thick graphene sheets 
with dislocation-network structures against a HOPG substrate. As for the former, the in-plane periodicity increased 
gradually in the direction of inclination, and it is easily changed by attachment of a nanographite flake on the 
nanographene sheets. The oscillation pattern can be explained by the interference of electron waves confined in the 
inclined nanographene sheets. As for the latter, patterns and their corrugation amplitudes depended on the bias voltage 
and on the terrace height from the HOPG substrate. The interference effect by the perturbed and unperturbed waves in the 
overlayer is responsible for the patterns whose local density of states varies in space.
\end{abstract}

\end{frontmatter}

\maketitle

\section{\label{sec:level1}Introduction}

A superperiodic pattern on metal surfaces has been observed on Cu(111) and Ag(111) surfaces by scanning tunneling 
microscopy (STM).\cite{ref1,ref2} This is interpreted as an interference wave pattern of free electrons that move on the Fermi 
surface originating from the surface and the bulk states, and are scattered by adatoms, step edges, and defects. Recently, 
a superperiodic pattern was also reported on semiconductor surfaces such as InAs/GaAs(111)\textit{A}.\cite{ref3} This pattern is also 
explained by interference in the two-dimensional (2D) electron gas, which are electrons generated due to the band bending 
caused by a surface reconstruction and confined in a very thin layer near the surface. As for 2D structures, layered 
materials such as graphite are other candidates in generating the electronic wave interference effect due to a large 2D 
anisotropy. As an indicative fact of a large anisotropy of graphite, the foliation is often generated by heat treatment at 
a high temperature or by the direct collision of a STM tip against a graphite surface.\cite{ref4} The foliated nanographene sheets 
are occasionally folded back after tearing a graphite surface, and in one case they are adhered to the graphite surface 
with its edge structure rolled like the carbon nanotube. In this case, the nanographene sheets except for a nanotube-like 
part interact with the graphite substrate and sometimes a superperiodic pattern appears as a consequence of a defaulted 
stacking. Actually, if the foliated nanographene sheets are not folded completely and not adhered tightly to the graphite 
substrate surface, the interaction between the nanographene sheets and the substrate becomes very week, in the 
nanographene, where the standing wave is expected to generate due to the electron confinement in the 2D structured 
nanographene. From its anisotropy, other types of superperiodic patterns are also generated by stacking faults. One is a 
moire pattern related to a relative rotation between adjacent layers and another is a pattern originating from the 
dislocation-network structures due to abrupt transitions of stacking with the lattice distortion in plane.\cite{ref5,ref6,ref7} Though 
the periodicity of a moire pattern and a pattern originating from the dislocation-network structures can be determined by 
the angle of the relative rotation and the periodicity of glides, respectively, the corrugation amplitudes observed by STM 
cannot be explained just by the calculated local density of states (LDOS) of different stacked graphite.\cite{ref5,ref8} Instead, 
the scattering potential at the interface between the overlayer (a faulted part) and the substrate should give rise to the 
interference of the perturbed and the unperturbed waves in the overlayer, which affects the corrugation amplitudes of 
patterns dependent on the bias voltage of STM and the overlayer height from the substrate.\cite{ref9}\\
\hspace*{10pt}In this paper, we report observed superperiodic patterns, whose periodicities vary gradually in plane, on nanographene 
sheets inclined with respect to a HOPG substrate, and superperiodic patterns originating from the dislocation-network 
structures, where the corrugation amplitudes depend on the bias voltage and the overlayer height from the substrate.

\section{\label{sec:level2}Electronic wave interference patterns\\ \hspace*{100pt}in nanographene sheets}

\begin{figure*}
\begin{center}
\includegraphics[width=14cm, clip]{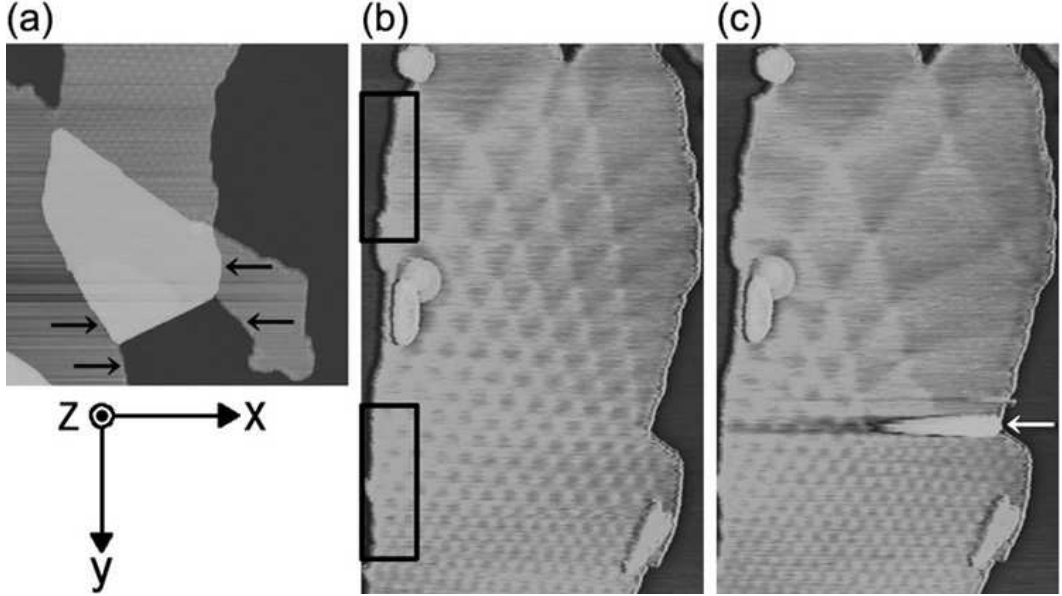}
\caption{\label{fig:ISIC1} (a),(b) A superperiodic pattern on nanographene sheets inclined with respect to the HOPG 
substrate observed by STM ($V_s$=0.2 V, $I$=0.7 nA). The image in (b) (400$\times$250 nm$^2$) is extended from the upper region of the 
image in (a) (400$\times$400 nm$^2$). Arrows in (a) indicate graphite edges generated by tearing the HOPG surface (400$\times$400 nm$^2$). 
The apparent sheets height from the substrate determined in (b) was $\sim$0.8 nm in the upper square and $\sim$0.9 nm in the lower 
square. (c) A different pattern formed by attachment of a nanographite flake on the nanographene sheets (400$\times$250 nm$^2$).}
\end{center}
\end{figure*}
A superperiodic pattern, whose periodicity changed gradually in plane, was observed by STM as shown in fig.1(a)-(b). The 
sample preparation is given elsewhere.\cite{ref5,ref11} The nanographene was fabricated by turning back after the 
heat-treatment-induced foliation at 1600 \symbol{"17}C, which is ascertained from the fact that the shape of a nanographene edge 
resembles that of a step edge of the substrate as indicated by arrows in fig.1(a). The apparent height of the nanographene 
sheets from the HOPG substrate was about 0.8 nm from a cross-sectional analysis of the upper square region in fig.1(b). The nanographene probably consists of a stacking of two graphene layers, which interact weakly with the HOPG substrate, by comparison with the interlayer distance of bulk graphite (0.335 nm). The height of the lower square region in fig.1(b) was about 0.9 nm, therefore, the nanographene inclined from the substrate with an average slope of $\Delta z/\Delta y$$\sim$$2\times 10^{-4}$. Note that the periodicity of the observed superperiodic pattern decreases in increasing the $y$ value. The corrugation amplitude of pattern also decreases in increasing the $y$ value from the cross-sectional analyses of the STM image in fig.1(b). Interestingly, the pattern in fig.1(b) changed into that in fig.1(c) by attachment of a nanographite flake on the nanographene sheets. By comparing two patterns in figs.1(b) and (c), it is evident that the periodicity changed remarkably at the upper part, though there was only a subtle change in the periodicity in the lower part. Such an effect on the patterns cannot be explained by a simple structural modulation. Instead, we can take the long-distance slope of nanographene sheets as the origin of the pattern and the interference effects of an electronic wave function due to a 2D structure of the nanographene interacting weakly with the substrate. In this case, the spatially varying periodicity can be explained by the electronic confinement effect in 2D-structured nanographnene with a potential gradient generated from the slope of the nanographene.
To reproduce the observed patterns, we will characterize the patterns theoretically on the basis of the interference effect in a free electron model at first.\cite{ref12} By assuming a linear potential $-Fy$ generated from the slope of the nanographene along the $y$ axis and a confinement effect in the $x$ axis due to the square-well potential within $-d/2$$<$$x$$<$$d/2$, where $d$ is the width of the nanographene sheets along the $x$ axis, the Schr\"{o}dinger equation is obtained:
\begin{equation}
[\frac{\hbar ^2}{2m}(\frac{\partial ^2}{\partial x^2}+\frac{\partial ^2}{\partial y^2})+V_x-Fy]\psi (x,y)=E\psi (x,y),
\end{equation}
where $V_x$ is a square-well potential with infinite depth. The solution in the well is a standing wave in the $x$ axis and an 
oscillatory wave that is given by the Airy function in the $y$ axis. The potential gradient constant $F$, which was derived 
from fitting the LDOS to the observed pattern shown in fig.1(b) in the $y$ axis, was estimated to be about $5\times 10^{-6}$ eV/nm, 
using the free electron mass. The potential variation for 200 nm along the $y$ axis was calculated to be about $1\times 10^{-3}$ eV, 
which is unrealistically small because it is one order of magnitude smaller than thermal energy $k_{\textrm{B}}T$ at room temperature, 
where $k_{\textrm{B}}T$ is the Boltzmann constant. This inconsistency comes from the difference of electronic structures between the free 
electron model and actual 2D graphite. Then we use the $k\cdot p$ perturbation model, taking the band structure of 2D graphite with the linear dispersion near 
the Fermi energy into consideration for making a more realistic model. The Hamiltonian around the $K$-point with the linear potential is given as,
\begin{equation}
H=\begin{pmatrix} -Fy & -i\gamma \frac{\partial }{\partial x}-\gamma \frac{\partial }{\partial y} \\ -i\gamma \frac{\partial }{\partial x}+\gamma \frac{\partial }{\partial y} & -Fy\end{pmatrix},
\end{equation}
where $\gamma $ is $(\sqrt{3}/2)a\gamma _0$, $a$ and $\gamma _0$(=2.7 eV) are the nearest-neighbor distance and the resonance integral between nearest-neighbor carbon 
atoms, respectively. From the Schr\"{o}dinger equation, $H\psi (x,y)$=$E\psi (x,y)$, the solution is a standing wave in the $x$ axis and 
sinusoidal wave of squared $y$ in the $y$ axis. The potential gradient constant $F$ is obtained as $5\times 10^{-3}$ eV/nm from the LDOS of 
eigen wave functions with the Hamiltonian around the $K$- and $K'$- points, where the variation of the potential over 200 nm 
along the $y$ axis is about 1 eV. This model satisfies conditions of the experimental results.

\section{\label{sec:level3}Electronic wave interference effec\\ \hspace*{70pt}in ultra thin graphene layers}

\begin{figure*}
\begin{center}
\includegraphics[width=16cm, clip]{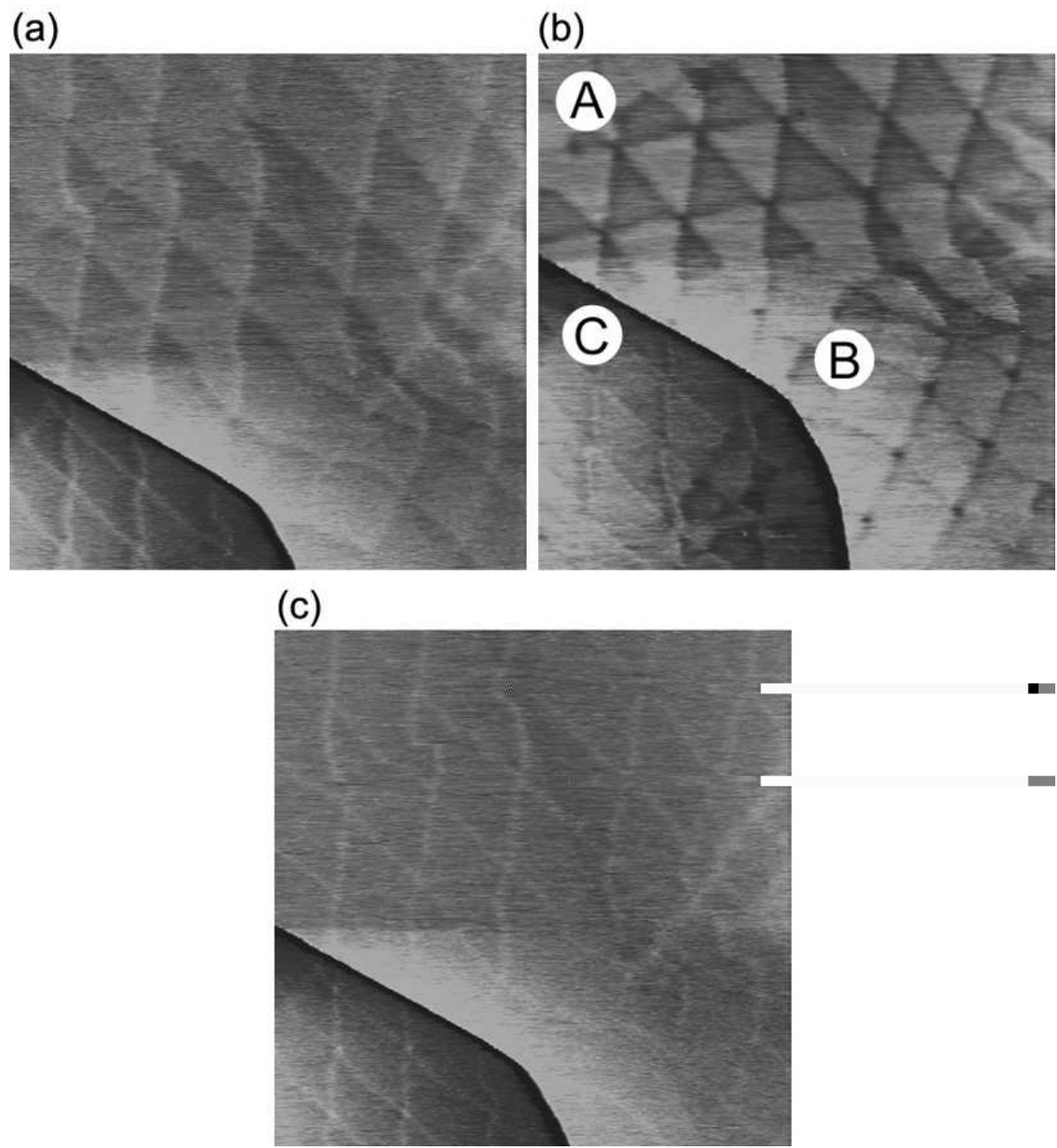}
\caption{\label{fig:ISIC2} Bias dependent STM images (500$\times$500 nm$^2$) of superperiodic patterns originating from a 
dislocation-network structure (constant-current mode; $I$=0.7 nA): (a)$V_s$=0.3 V, (b)$V_s$=0.02 V, (c)$V_s$=0.5 V, respectively. 
A line that runs in bottom left part is a step edge. The apparent height of lower terrace (at lower left of the figure) 
and upper terraces (at center) from the substrate correspond to two and three graphene layers, respectively.}
\end{center}
\end{figure*}
Superperiodic patterns that extended over several $\mu $m$^2$ were observed by STM on ultra thin graphite layers. Figure 2(a) shows 
a part of those patterns in an 500$\times$500 nm$^2$ image near a step edge observed at a sample bias voltage of 0.3 V. In this image, 
the patterns are sharp-shaped and the periodicity is very large ($\sim$100 nm). We consider that the patterns come from the 
dislocation-network structures due to the similarity of the patterns to those reported by STM and transmission electron 
microscopy observations.\cite{ref6,ref7} Cross-sectional analyses of STM images indicate that apparent heights of the lower and upper 
terraces in fig.2 correspond to two and three graphene layers from the substrate, respectively. Figure 2(b) shows an image 
that is shifted to bottom direction of fig.2(a) by 200 nm close to the Fermi energy ($V_s$=0.02 V). In this image, we can 
classify patterns into three shapes. Triangular-, rhombic- and net-shaped patterns were observed in domains A, B and C, 
respectively. The geometric patterns in domains B and C appears to have an inversed contrast though those patterns come 
from an array of dislocations as is evident from the fact that the patterns are continuous at the step edge. The corrugation 
amplitude of the patterns in domain A and B in fig.2(b) is larger than that in fig.2(a). In contrast, the corrugation 
amplitude of the patterns in domain C in fig.2(b) is smaller than that in fig.2(a). These phenomena cannot be explained 
just by the LDOS of stacking faulted parts and the tunneling probability dependent on a gap between a tip and a sample. By 
increasing the bias voltage, the patterns on the upper terrace in fig.2(b) changed into the net-shaped pattern although the 
pattern shape on the lower terrace does not change, as shown in fig.2(c). This phenomenon cannot be explained by a change 
of the dislocation-network structures driven by the applied bias voltages because the periodicity did not change. We can 
consistently explain those phenomena as the interference effect originating from the perturbed and unperturbed waves in the 
overlayer, by assuming the scattering potential at the interface to generate the patterns.\\
\hspace*{10pt}Here, we theoretically treat the interference effect for explaining the bias dependence of superperiodic patterns by a free 
electron model assuming an effective mass for electrons.\cite{ref9,ref11} An in-plane square-patterned potential with a periodicity 
of 2$L$ at the interface is employed for a calculation of the LDOS of the wave function confined in the plane to generate an 
abrupt potential change associated with the dislocation-network structures. Although the square-patterned potential is not 
consistent with the experimentally observed triangular-shaped pattern, the model makes the analysis considerably simple 
without any loss of validity. We place a square potential with $L$/3 in width and 4$v_0\delta(z)$ in height at the lines dividing 
the pattern into individual geometric unit as a simple model for reproducing the patterns in domain B and C. By the Fourier 
analysis, this potential is expressed as,
\begin{eqnarray}
V(x,y,z)=(\hbar ^2/m_\bot )v_0 \sum_{n}a_n\delta (z)\nonumber \\ \hspace{15pt}\times (e^{iq_{xn}\cdot x}+e^{-iq_{xn}\cdot x}+e^{iq_{yn}\cdot y}+e^{-iq_{yn}\cdot y}),
\end{eqnarray}
where $\hbar $ is the Planck constant over 2$\pi $, $m_\bot $ is the effective mass along the $z$ axis, $a_n$ is the $n$th component which equals to 
$\{2/(n\pi )\}\{\sin(n\pi )-\sin(5n\pi /6)\}$ for the square potential, and $q_{xn}$ and $q_{yn}$, which take discrete values 
$(n\pi /L)$($n$=1,2,...), are the $n$th wave vectors along the $x$ and $y$ axes, respectively. We can find a wave function in 
the form of
\begin{equation}
\Psi (x,y,z)=\sum_{q_x,q_y} A_{q_x,q_y}(z)e^{i(q_x\cdot x+q_y\cdot y)},
\end{equation}
where $A_q(z)$ is a wave function for the $z$ component. The perturbed wave is obtained as a solution of the Schr\"{o}dinger equation 
with the discontinuity of $d^2ĵ/dz^2$ at the interface. Provided a perturbed wave number and a unperturbed wave number are $k'$ 
and $k$, respectively, the LDOS of a superperiodic pattern at the surface is roughly propotional to the sum of 
$[-a_nk\sin(kl)\cos(k'l)\{\cos(q_{xn}\cdot x) +\cos(q_{yn}\cdot y)\} ]$ with index $n$, where $l$ is the overlayer height from the interface. The effective mass of electrons was 
estimated from the band dispersion of graphite. The calculated LDOS suggested that the corrugation amplitude of the superperiodic pattern could increase, decrease or invert, 
dependent on the overlayer thickness and the bias voltage, and consistently reproduced the experimental results.

\section{\label{sec:level4}Summary}

A superperiodic pattern, whose periodicity changes in plane, was observed by STM on nanographene sheets that inclined 
from the HOPG substrate with weak interaction from the substrate. The periodicity changed remarkably by attachment of a 
nanographite flake on the nanographene sheets. The pattern can be explained as the interference effect of an electronic 
wave in the nanographene sheets on the basis of the $k\cdot p$ perturbation model that assumes a linear potential in the direction of slope and the electron 
confinement effect.\\
\hspace*{10pt}Superperiodic patterns, which come from the dislocation-network structure, were also observed by STM. Individual shapes of the observed patterns were different at 
terraces with different number of layers from the substrate and their contrasts changed depending on the bias voltage. The patterns can be explained in terms of the 
interference effect by perturbed and unperturbed waves in the ultra thin graphite overlayer.

\begin{ack}

The authors are grateful to Prof. Katsuyoshi Kobayashi (Ochanomizu University), for fruitful discussion. They also thank to Dr. A. Moore for his 
generous supply of HOPG sample. The present work was supported partly by the Grant-in-Aid for `Research for the future' 
Program, Nano-Carbon, and 15105005 from Japan Society for the promotion of Science. One of the authors (K.H.) acknowledges 
the financial support from NEDO via Synthetic Nano Functional Materials Project, AIST, Japan.
\end{ack}

\end{document}